\documentstyle[aas2pp4]{article}

\def\B{{\bf B}}

\def\eset{e_{\rm s}/e_{\rm k}}

\def\Fc{\bf F_{\rm c}}
\def\Fs{\bf F_{\rm s}}
\def\Gammad{\Gamma_{\rm d}}
\def\game{\gamma_{\rm e}}
\def\gameff{\gamma_{\rm e}}
\def\Gammas{\Gamma_{\rm s}}
\def\kfor{k_{\rm for}}
\def\o{{\rm o}}
\def\ppm{{\it ppm}}
\def\rc{r_{\rm c}}
\def\rhoic{\rho_{\rm ic}}
\def\u{{\bf u}}
\def\wp{{\bf \omega}_{\rm p}}

\slugcomment{Submitted to {\it The Astrophysical Journal}}

\begin{document}

\title{Influence of cooling-induced compressibility on the structure of
turbulent flows and gravitational collapse}

   \author{Enrique V\'azquez-Semadeni,}
    \affil{Instituto de Astronom\'\i a, UNAM, Apdo. Postal 70-264, M\'exico, D.\
F.\ 04510, M\'exico}
\affil{e-mail: {\tt enro@astroscu.unam.mx}}
\author{Thierry Passot and Annick Pouquet}
\affil{Observatoire de la C\^ote d'Azur, B.P.\ 229, 06304, Nice
             Cedex 4, France}
\affil{e-mail: {\tt passot,pouquet@obs-nice.fr}}

\begin{abstract}
We investigate the properties of highly
compressible turbulence, the compressibility arising from a small
effective polytropic exponent $\game$ 
due to cooling. In the limit of small $\game$, the density jump at
shocks is shown to be of the order of $e^{M^2}$, 
much larger than the $M^2$ jump
associated with high-Mach number flows in the isothermal regime.
In the absence of self-gravity,
the density structures arising in the moderately compressible case consist
mostly of patches separated by shocks and behaving like waves,
while in the highly compressible case clearly defined long-lived 
object-like clouds emerge. The transition from wave-like to
object-like behavior 
requires a change in the relative phase of the density and velocity
fields analogous to that in the development of an instability.
When the forcing in the momentum equation is purely compressible, 
the rotational energy decays monotonically in time,
indicating that the vortex-stretching term is not efficient in
transferring energy to rotational modes. 
This property may be at the origin of the low amount of rotation found
in interstellar clouds. Vorticity
production is found to rely heavily on the presence of additional terms in the
equations, such as the Coriolis force at large scales and the Lorentz force at
small scales in the interstellar medium, or on the presence of
local sources of heating.
In the presence of self-gravity, we suggest that turbulence can produce bound
structures for $\game < 2(1-1/n)$, where $n$ is the typical
dimensionality of the turbulent compressions. 
We support this result by means of numerical simulations
in which, for sufficiently small $\game$,
small-scale turbulent density fluctuations eventually
collapse even though the medium is globally stable. This result is
preserved in the presence of a magnetic field for supercritical
mass-to-flux ratios.
At larger polytropic exponents, turbulence alone is not
capable of producing bound structures, and collapse can only occur when 
the medium is globally unstable.
This mechanism is a plausible candidate for the
differentiation between primordial and present-day stellar-cluster
formation and for the low efficiency of star formation.
Finally, we discuss models of the interstellar medium at the kpc scale
including rotation, which restores a high-$\game$ behavior. 

      \keywords{ISM: clouds -- instabilities -- magnetohydrodynamics
        -- turbulence -- Cosmology: miscellaneous}
   \end{abstract}

%

\section{Introduction}

Astrophysical flows are expected to be highly turbulent, with a wide
variety of mostly compressible energy sources at various scales.
Thus we can expect the nature of astrophysical turbulence to be
quite different from standard laboratory turbulence, which is mainly a
vortical phenomenon. If the flow is subject to strong heating
processes and radiative cooling, with thermal time scales 
substantially shorter than the dynamical ones (\cite{spit50}), as is
the case in the interstellar medium (ISM) or the post-recombination
medium in cosmology, 
rapid thermal
equilibrium can be established, and the fluid behaves as a polytropic gas
in which $P \propto \rho^{\gameff}$, where $\gameff$ is an
equivalent polytropic exponent. Such systems have been discussed 
both in the linear (\cite{elm91}) and nonlinear regimes
(V\'azquez-Semadeni et al.\ 1995, hereafter referred to as
\cite{paperI}; Passot et al.\ 1995, hereafter \cite{paperII}). 
In particular, in the models of the ISM at the 10--1000 pc scales we have
presented in previous work, the thermal time scales range from
$10^{-1}$ to $10^{-4}$ times the dynamical timescales (\cite{paperI}; see
also \cite{elm93}).
For standard cooling functions
(e.g., \cite{dalg}), which depend on the metallicity and ionization fraction
among other factors, $\gameff$ can take values smaller than one and even
negative (leading to the isobaric 
mode of the thermal instability). As $\gameff$
approaches zero, pressure gradients are almost absent, leading to a
rather inelastic behavior with a 
very high compressibility, a behavior which is
reminiscent of Burgers flows even when the Mach number based on the
isothermal sound speed (the quantity normally reported in 
observational studies) is of order unity.

In this paper we investigate the influence of the degree of
compressibility of the flow (measured by either the effective polytropic
index $\gameff$ or the actual Mach number $M_a$) on both the properties of
the turbulence and the mechanisms of formation of 
density structures. It is important to predict whether turbulence is
dominated by vortical or compressible motions, depending on 
the type of forcing imposed on the flow,
and the additional physical processes considered.  In particular, it
is of interest to investigate the nature of the transfer between
compressible and rotational modes of the velocity field. This problem
has been investigated in the low-to-moderate compressibility regimes
by Kida \& Orszag (1990a,b), who found that transfer occurs mostly
from rotational to compressible modes, and only in an efficient manner
above $M_a \sim 0.3$ (see also \cite{pp87}). 
In the present paper we extend their
results to the highly compressible regime, and use two- and three-dimensional
numerical simulations to show that, for purely
compressible forcing (such as that originating from stellar winds),
vorticity decreases steadily. We also discuss alternative vorticity
production mechanisms, such as the baroclinic term or the
presence of shear, the Coriolis
force, and magnetic fields. Throughout the paper, we address the discussion 
mostly towards processes in the interstellar medium at various scales,
although we consider generalizations which should be applicable to a
wider variety of astrophysical flows, in particular in Cosmology and
in molecular clouds.

In the context of the ISM, both the amplitude and
persistence of density fluctuations
are highly dependent on the degree of compressibility of the flow, and thus the
latter can affect the cloud and star formation mechanisms. 
In the linear regime,
Elmegreen (1989) has proposed that the gravitationally unstable scales
are greatly reduced in the presence of dissipation through cloud
collisions, while \cite{ren93} have studied the
modifications to the Jeans length in the vicinity of the thermal
instability. In addition, various workers have discussed the formation of 
self-gravitating clouds induced by external velocity fields
(e.g., \cite{hunfle82}; \cite{hun86}; \cite{elm93};
\cite{paperI}; \cite{padoan}), although without consideration of the
direct dependence on
an effective polytropic index $\gamma_e$.
In this paper we also discuss, based on numerical simulations, 
the ability of turbulence to form
gravitationally bound structures which are the sites of star
formation, as well as their characteristic scales, in relation to
$\gameff$. This problem will
be addressed in the context of both a polytropic gas (either with a
single value of $\gameff$ or a piecewise index depending on the
density) and a fully thermodynamic model of the ISM including parameterized
heating and cooling, as introduced in \cite{paperI} and, with rotation and
magnetic field (\cite{paperII}). Note that in this paper we only consider 
the large-scale effects
of turbulence in the formation of density structures, neglecting its
well known effects for cloud support at small scales (e.g., \cite{chandra};
\cite{bonaz}; \cite{leorat}; \cite{vsgaz}; see also the review by 
\cite{scalo87} and references therein). However, our results should
also be applicable to a cloud-fluid in which the pressure includes a
sub-grid turbulent component (e.g., \cite{hunfle82}; 
\cite{ssm84}; \cite{elm91});
sub-grid cloud collisions provide an effective power-law ``cooling''
mechanism (e.g., \cite{elm91}).

The outline of the paper is as follows.
In the next section we describe the numerical approach and define the
various models. Section
\ref{hicomp} is devoted to the influence of
various degrees of compressibility on randomly forced compressible
turbulence, when the forcing is only in the velocity
equation and is only potential, and to 
the production of vorticity in the highly
compressible regime by the vortex-stretching term and other mechanisms, in
two and three dimensions. The next Section then deals with the
implications for scale-free astrophysical flows when self-gravity and
magnetic fields are included. Section \ref{allinc} considers the results
of the previous sections in the context of models of the ISM at the kpc scale 
based on the one presented in \cite{paperII}. 
Finally, \S\ \ref{concl} is the conclusion.

\section{The models} \label{models}

In this paper we use several sets of hydrodynamic equations to simulate
highly compressible flows, ranging from a simple barotropic flow to
the full model of the ISM used in \cite{paperII} (with the addition of winds 
or a random forcing $\Fc$
in the velocity equation). The latter reduces to
the following equations, which represent the ISM on the plane of the
galactic disk:
\begin{equation}
{\partial\rho\over\partial t} + {\nabla}\cdot (\rho\u) = \mu
\nabla^2 \rho, \label{cont}
\end{equation}
\begin{eqnarray}
{\partial\u\over\partial t} + \u\cdot\nabla\u =
-{\nabla P\over \rho} 
- \Bigl({J \over M_a}\Bigr)^2 \nabla \phi +\nonumber \\
{1 \over \rho}
\bigl(\nabla \times {\bf B}\bigr) \times {\bf B}
 - 2 \Omega \times \u - {\nu_8} {\nabla^8\u}+\nonumber \\
\nu_2 (\nabla^2 \u + \frac{1}{3} \nabla \nabla \cdot \u)
+ \Fs + \Fc,\label{mom}
\end{eqnarray}
\begin{eqnarray}
{\partial e\over\partial t} + \u\cdot\nabla e = -(\gamma -1)
e\nabla\cdot \u + {\kappa}_T {\nabla^2e\over \rho} + \nonumber\\
\Gammad + \Gammas - \rho \Lambda, \label{ener}
\end{eqnarray}
\begin{equation}
{\partial {\bf B}\over\partial t} = \nabla \times (\u \times {\bf B})
- {\nu_8} {\nabla^8{\bf B}} + \eta \nabla^2 {\bf B}, \label{magn}
\end{equation}
\begin{equation}
\nabla^2 \phi=\rho -1, \label{poisson}
\end{equation}
\begin{equation}
P=(\gamma-1)\rho e, \label{eqstat}
\end{equation}
\begin{equation}
\Gammad({\bf x},t)={\Gamma_{\o}(\rho/\rhoic)^{-\alpha}}, \label{gammad}
\end{equation}
\begin{equation}
\Gammas({\bf x},t) = 
\cases{{\rm constant}   & if $\rho({\bf x},t_0) >
        \rho_{\rm cr}$ \cr
        & and $0 < t-t_0 < \Delta t_s$\cr
        0               & otherwise\cr}, \label{gammas}
\end{equation}
and
\begin{equation}
\Lambda = \Lambda_i T^{\beta_i} \ \ \ {\rm for}\ \  T_i \le T < T_{i+1},
\label{lamda}
\end{equation}
where

{\begin{tabular}{llll}
$T_1 = 100$     & $\Lambda_1= 1.14 \times 10^{15}$ &$\beta_1=2$\\
$T_2 = 2000$    & $\Lambda_2= 5.08 \times 10^{16}$ &$\beta_2=1.5$\\
$T_3 = 8000$    & $\Lambda_3= 2.35 \times 10^{11}$ &$\beta_3=2.867$\\
$T_4 = 10^5$    & $\Lambda_4= 9.03 \times 10^{28}$ &$\beta_4=-0.65$\\
$T_5 = 4 \times 10^7$.
\end{tabular}}

\noindent
The temperature $T$ is measured in $K$ and $\Lambda$ in erg ${\rm
  s}^{-1}$ ${\rm g}^{-2}$ ${\rm cm}^3$. 
Note that the stellar heating $\Gammas$ is spread out on a few pixels
around a stellar center by a convolution with a Gaussian profile. In eq.\
(\ref{gammas}), $t_0$ is the time at which the density threshold
$\rho_{\rm cr}$ is reached at point {\bf x} (see Paper I and II for further 
details).

As usual, $\rho$ is the density, {\bf u} is the fluid velocity, $e$ is the
internal energy per unit mass, $P$ is the thermal pressure,
\B\ is the magnetic induction,
$\Omega$ is the angular velocity of the rotation,
and $\phi$ is the gravitational potential. 
The ratio $\gamma = c_p/c_v$ of specific heats at constant pressure
and volume respectively is $\gamma=5/3$. The temperature is related
to the internal energy by $e =c_vT$. The nondimensionalization is the
same as in \cite{paperII}; the resulting nondimensional parameters
are $J$, the number of Jeans lengths in the integration box, whose
side is $2 \pi$, and
$M_a$, the Mach number corresponding to the characteristic velocity
$u_0$ at the initial mean temperature of the flow (note that we take
$c_{\rm v}=[\gamma (\gamma-1) M_a^2]^{-1}$). The magnetic
field is written as $\B =B_0 {\bf e}_x+{\bf b}$ where $B_0$
represents the uniform azimutal component of the field, and ${\bf
  b}$ a superimposed fluctuating component. The shearing forcing
$\Fs$ is accomplished by fixing one component of the largest Fourier
mode in order to impose a velocity profile $u_x=-0.375 \sin y$; the additional
term $\Fc$ is discussed further below.

These equations are solved both in two and three dimensions with
periodic boundary conditions, using a Fourier pseudospectral
technique with a resolution of $128^2$ collocation points 
for two-dimensional runs and $64^3$ for the
three-dimensional ones. A hyperviscosity scheme with
a $\nabla^8$ operator is used, which confines viscous
effects to the smallest resolved scales and allows for the use of
much smaller values of the kinematic viscosity $\nu_2$ and magnetic diffusivity
$\eta$ than what would be possible when $\nu_8 =0$. The adjunction of
the second-order viscosity and diffusivity, compared to the model used in
\cite{paperI} and \cite{paperII}, has been made necessary in the highly
compressible regime in order to ``filter'' out the oscillations in the
vicinity of strong shocks (\cite{ppjcp}). In addition, 
a mass diffusion $\mu\nabla^2\rho$ is added to the continuity equation
in order to smooth out the density gradients, thus allowing the simulations to
reach higher r.m.s.\ Mach numbers. The values of the various
coefficients for the $128^2$ runs 
are $\nu_8=8\times 10^{-12}$, $\nu_2=\eta=2\times
10^{-3}$, $\mu=3\times 10^{-2}$ and $M_a=1$ throughout the
paper. For the 3D runs 
we take  $\nu_8=10^{-9}$, $\nu_2=2.5\times 10^{-3}$, and $\mu=4 \times
10^{-2}$.
We refer the reader to \cite{paperII}
for further details concerning the technique and the model terms.
The additional acceleration $\Fc$ introduced in the velocity
equation has been taken either as a random vector field
with a $k^{-4}$ power-law spectrum peaked at a wavenumber $\kfor$, white noise in time
and with an adjustable ratio $\rc$ of compressible to solenoidal
components, or as a wind-mimicking momentum source of the form $\Fc= -
\nabla \psi$, where $\psi$ is proportional to the stellar heating
$\Gammas$. Finally, the
initial conditions for all variables are Gaussian fluctuations with random 
phases. For the velocity field, the fraction of solenoidal to compressible
energy is an additional adjustable parameter.

According to the results of \cite{paperI} and \cite{paperII}, due to
the short thermal time scales, the flow behaves essentially as a piecewise
polytropic gas, except in the regions of star formation. As a
consequence, we also solve a simplified set of equations (hereafter
denoted \ppm, for Piecewise Polytropic Model) in which we eliminate
the internal energy equation and use a piecewise polytropic equation
of state given by
\begin{equation}
P=a \rho^{{\gameff}_{i}} \equiv
\Bigl[{\Gamma_\o \rhoic^\alpha \over \Lambda_{\rm i}}\Bigr]
^{1/\beta_{\rm i}}{\rho^{{\gameff}_{i}}\over \gamma M_a^2} \ \ , \label{pppm}
\end{equation}
with $i=1,...,5$, in conjunction with the five temperature domains of 
eq. (\ref{lamda}). Note that in
the barotropic case the nondimensionalization of the equations leads
to an explicit factor of $M_a^{-2}$ in the pressure, the sound speed
in the equilibrium state of temperature $T_{\rm eq}$
being given by $c_{\rm e}^2=(\game/\gamma M_a^2)T_{\rm eq}$. 
The effective polytropic index is given, in terms of the exponents $\beta_i$
in (\ref{lamda}) and $\alpha$ in eq.\ (\ref{gammad}), by:
\begin{equation}
{\gameff}_{i}= 1-{(1+\alpha) / \beta_i} \ \ .
\label{albega}
\end{equation}
Note that ${\gameff}_{i}$ was noted $\gamma_{eff}$ in Paper II, but the present
notation stresses its dependence on the interval of temperature in the
cooling function. Finally, the boundaries of
the different temperature ranges in the cooling function for the full
model correspond to densities given by
\begin{equation}
\rho_i=\bigl({\Gamma_\o \rhoic^\alpha \over \Lambda_{\rm
i}}\bigr)^{\frac{1}{1+\alpha}}T_{i+1}^{-\frac{\beta_i}{1+\alpha}}.
\end{equation}

For convenience, we adjust the degree of compressibility of the flow
(the value of ${\gameff}_{i}$) by varying the exponent $\alpha$ of the diffuse
heating. This is equivalent to varying the exponents $\beta_i$ in 
all intervals of the cooling function. Although this is not a very
realistic approach, it provides us with the simplest means for varying
the flow's ``hardness'', i.e. the flow's resistance to
compression. Actual modifications to the cooling function 
will be discussed in a future paper (\cite{franco}).

A further level of simplification, useful to single out purely fluid
dynamical properties of supersonic turbulence, consists of using a
single polytropic exponent $\game$. In this case, the pressure is
taken as $P=(\game M_a^2)^{-1} \rho^{\game}$, leading to a sound
speed $c_{\rm e}=(1/M_a)$ in the equilibrium state $\rho=1$.
In the following, we refer to 
these equations as the B-model, for Barotropic model.

Throughout the paper, we will refer to a rather large number of
simulations, whose most relevant parameters are summarized in Table 1.

\section{Some properties of highly compressible flows}
\label{hicomp}

\subsection{Density structures} \label{denstruct}

As discussed in the introduction, low values of $\gameff$ lead to very
high compressibilities, and in fact the degree of compression
reached in this case is much higher than the one obtained by increasing
the Mach number. This can be seen by calculating the density jump
$X\equiv \rho_2/\rho_1$ 
across a shock in a barotropic gas of index $\gameff$. It is easy to
show that $X$ satisfies 
\begin{equation}
X^{1+\game} - (1+ \game M^2)X + \game M^2=0,
\end{equation}
where $M$ is the Mach number upstream of the shock. From this
equation, we recover the fact that the compression ratio for an
isothermal shock ($\game =1$) is $M^2$, but we also see that, for
$\game \ll 1$, $X$ is approximately given by $X \sim (1 + \game
M^2)^{1/\game}$. For a fixed value of $M$, this expression tends to
$e^{M^2}$ as $\game \rightarrow 0$, a density jump which can be much
larger than the isothermal one.

We now discuss the nature of the density condensations
present in the flow as observed in our simulations,
and their relation to the value of the polytropic
exponent. 
For simplicity, we consider  the B-model
and the \ppm\ equations, with a purely compressible random forcing in the 
velocity equation.

Even though the boundary between moderate and high compressibility
is not sharp, the two regimes lead to quite different density
structures. In the former case, the density field consists of patches
separated by shocks, and whose lifetime is the sound crossing time
within the patches. This is illustrated in fig.\ \ref{hilowgam}
(left), where we 
display a typical snapshot of the density field in a simulation
labeled run 63, with $\game =
1.5$ and $\kfor=4$. Note
that the highest peaks form at the intersection of shocks. Such a
behavior is analogous to what is observed in decaying flows, both in
2D (\cite{pas88}) and in 3D (\cite{ppw3d}). 

As $\game$ is lowered, the
density peaks become higher and narrower, and have longer lifetimes since
the effective sound speed decreases. This is illustrated in 
fig.\ \ref{hilowgam} (right), showing the density field of run 65,
with $\game = 0.3$.  
In fact, when using the \ppm\
equations with the fiducial values of the parameters used in Paper II
($\alpha =0.5$),
$\gameff$ can locally reach zero for intermediate densities,
and we then observe a transition of behavior from wave-like to
object-like. We exemplify this behavior in fig.\ \ref{ppmhilow},
which shows the density field of two two-dimensional
simulations using the \ppm\ equations, the first one with
$\alpha=-0.5$ (run 54, left), and the second with $\alpha=0.5$ (run 53, right).

In order to understand the distinction between wave-like and
object-like behavior, consider a
passive scalar, which is defined as a quantity $s$ obeying the pure
advection equation $\partial s
/ \partial t = -\u \cdot \nabla s$, and consider its motion relative to a
density perturbation. We define an object as a density perturbation
within which the passive scalar remains for a substantial period of
time, whereas in a wave the passive scalar and the density
fluctuations drift apart. For example, a one-dimensional 
wave packet $\rho=\rho(x-u_0 t)$,  moves
as a whole at speed $u_0$. 
Adams et al. (1994)  have shown that, in the context of a logatropic
equation of state of the form $P\sim c_{\rm e}^2\rho+p_0 \log (\rho/\rho_R)$
%
%
such wave packets can behave as solitary waves.
Conservation of mass leads to
$(u-u_0)\rho=cst$, showing that the fluctuations of $\rho$ and $u$
are ``in phase'' (fig.\ \ref{rhov phases}; see also fig.\ 2 of \cite{paperII}).
The scalar $s$, 
advected at the flow velocity $u\not= u_0$, will as a consequence
drift with respect to the density perturbation. On the other hand, in the case 
of an object (i.e., a density clump), the velocity field is either convergent 
towards its center (``out of phase''  with the density perturbation), or is
zero inside and contains two accretion shocks at its edges so that
the passive scalar remains within the boundaries of the density
fluctuation.
%
%
%
The transition from wave to object can spontaneously happen
when the flow becomes unstable, e.g.
thermally or gravitationally. In the presence of turbulence this behavior is
continuously approximated as the polytropic exponent approaches
zero. In this case, the flow behaves almost as a Burgers flow, with
large density peaks at shock locations instead of the density 
jumps appearing in flows with finite pressure.
Therefore, the clouds have long lifetimes
and are only affected by collisions with other clouds. Note that in
the limiting case of zero pressure gradient, cloud (shock)
collisions become
entirely coalescing (``sticky''), since disruption requires
the presence of vorticity, whose dynamics is discussed in the next
section.

\subsection{Evolution of vorticity in highly compressible turbulence}

\subsubsection{The role of vortex stretching} \label{vortstret}

It is first
useful to write the equations for the potential vorticity $\wp
\equiv {\bf \omega} / \rho = (\nabla \times \u)/\rho$ 
and the dilatation $\nabla \cdot \u$ in the
barotropic case. Neglecting the dissipation terms proportional to $\nu_8$ and 
$\nu_2$, they read
\begin{equation}
{\partial \wp \over \partial t} + \u \cdot \nabla \wp = \wp \cdot
\nabla \u \label{vorteq}
\end{equation}
\begin{equation}
{\partial \nabla \cdot \u \over \partial t} + \nabla^2 {u^2 \over 2} -
\omega^2 + \u \cdot \nabla \times \omega + {a \game \over \game -1}
\nabla^2 \rho^{\game - 1} = \nabla \cdot \Fc.
\end{equation}
Note that these equations are asymmetrical: dilatation can
easily be produced out of rotational motions (from the $- \omega^2$
term), whereas vorticity production can only originate from the
vortex-stretching term $\wp \cdot \nabla \u$ in three dimensions (3D) (see also
\cite{kornsca96}.
In the two-dimensional inviscid case, 
it is well known that there is conservation
of any functional of the potential vorticity (e.g., \cite{pedlovsky}) in the
absence of shocks. When oblique shocks are present, the vorticity
contained in them will be
dissipated much more rapidly than elsewhere in the flow in the
presence of viscosity.
Moreover, since $\wp$ obeys a passive scalar equation (in 2D), the vorticity
${\bf \omega}$ (which is larger in dense regions) will tend to be
confined inside clouds or at their edges in the highly compressible
regime, in which, as mentioned in \S \ref{denstruct}, clouds behave as
objects. This was observed in the simulations of the full model with
$\alpha=0.5$ (\cite{paperII}), but not in those of \cite{paperI}
with $\alpha=0$, {\it i.e.} with a higher $\game$.

We now examine whether the vortex-stretching term 
can contribute significantly to the transfer
of energy from the compressible to the solenoidal modes of the
velocity. Note that, even though in 3D the presence of the vortex
stretching term does not allow us to draw any conclusion regarding the
conservation of $\wp$, it is nevertheless relevant to consider the
equation for this quantity, since it avoids the need to take into
account the compression term $- \omega \nabla \cdot \u$ in the
vorticity equation, which only reflects mass conservation.

Kida \& Orszag (1990a) have investigated the transfer
between internal modes and compressible and rotational kinetic modes
in three dimensional, forced, compressible turbulence at moderate Mach
numbers. They find a negligible transfer between the compressible and
rotational kinetic modes. Here we extend this analysis to highly compressible
flows, using the \ppm\ equations. Starting with 100\% solenoidal
motions, and stirring the flow with a purely potential acceleration,
we observe, for a wide range of values of $\gameff$, that the ratio of
rotational to total kinetic energy per unit mass
\footnote{Note that the integrated energies per unit volume and
  per unit mass typically differ by no more than a few percent.}
$\eset$ always decays, as shown in
fig.\ \ref{esetlowhi}.  Together with other runs discussed below in \S 3.3.1,
this figure shows the evolution of $\eset$ for two 3D
runs at resolution $64^3$ solving the \ppm\ equations, one with
$\alpha=0.5$ (run 60), and the other with $\alpha=-0.5$ (run 67). 
The latter is thus ``harder''. Note also that, 
after the initial transients have
subsided, the rate of decay appears to be independent of $\game$, at
least for the range of values considered here. 
Thus, the stretching term does not appear to
contribute to energy transfer from the compressible to the rotational
modes.
 
The above result is rather unexpected, since vortex stretching is
the main mechanism for vorticity production
in incompressible turbulence. Instead, for highly compressible
flows, vortex stretching appears to contribute mostly to the destruction of
potential vorticity. We suggest the following mechanism. Let us first
consider the case in which potential vorticity is a small
perturbation on an irrotational flow. Since eq.\ (\ref{vorteq})
is identical to that for the magnetic
field, it is legitimate, in analogy with the kinematic dynamo
mechanism, to investigate whether $\wp$ can increase
on scales large compared to that of the potential flow, assumed
given. As is well known (e.g., \cite{krause}), the growth of the
magnetic field in a slow dynamo happens through the presence of helicity in
the flow. In the present case, however, the helicity of the basic flow is
identically zero, rendering the growth of potential vorticity
unlikely at dominant order. Nevertheless, we must consider the
possible amplification of $\wp$ on a scale comparable to that of the
basic flow (as in the context of the fast dynamo problem). In order to
investigate this problem, a knowledge of the structure of the basic
flow is necessary. As discussed in the previous section (see fig.\
\ref{hilowgam}), the  
highly compressible regime is dominated by the presence of locally
two-dimensional sawtooth shocks (see also \cite{kadom}). As already
observed by Kida \& Orszag (1990b), vortex lines tend to be
antiparallel with density (and velocity)
gradients. A possible explanation of this observation is that, on both
sides of a shock, the velocity gradient matrix has positive
eigenvalues, which tend to amplify vortex fluctuations in the
direction perpendicular to the shock, while the vortex stretching term
would have no effect on vortex lines orthogonal to the density
gradient. However, while the vorticity is amplified in these expansion
regions, it is also advected towards the shock from both sides, where
it is then destroyed since the velocity gradient matrix has negative
eigenvalues in these regions of strong compression.  In summary, in
highly compressible regimes, shocks are expected to
act like sinks in which potential vorticity is drained from the flow. 

In the fully nonlinear
case where rotational modes are initially of the same order of magnitude
as compressive ones, the numerical experiments shown in fig.\ \ref{esetlowhi}
indicate that the decay of rotational
energy persists, suggesting that the above mechanism is still at play.

\subsubsection{The baroclinic term}\label{baroclinic}

Another important source of vorticity in the fully thermodynamic case
is the so-called baroclinic
term, $\nabla P \times \nabla \rho / \rho^2$, 
which is known to be active behind curved shocks or at their
collisions (\cite{pp87}; \cite{fleck}; \cite{klein94}). To test its
effect, we have performed a new simulation, run 68, identical to run
60 described above, but using the
full thermodynamic equations. We find that the
rotational modes still decrease steadily in time, as also shown in fig.\
\ref{esetlowhi}. This indicates that the
baroclinic term has a globally negligible contribution to the
vorticity production, being overwhelmed by the draining effect of the
vortex-stretching term in negative dilatation regions. This remark is
corroborated  by the observation made by \cite{ppw95} that on 
average the
baroclinic term is $\sim 20$ times smaller than the stretching term in
compressible flows forced by an external shear. Note that, behind the
intersection of shocks, vorticity creation can be locally important
(\cite{pp87}), whereas in other regions the pressure and density
gradients are always close to parallel (\cite{kior90b}).

In fact, vorticity production behind shocks does not require the contribution
of the baroclinic term (\cite{hayes57}), the production mechanism relying on
simple geometrical considerations. In the present simulations such
vorticity production is not seen, possibly due to the lack of
resolution or to the fact that the shocks that form remain straight
(occurence of such vorticity production has been observed in high-resolution
simulations of collisions of strong SN shock waves with spherical 
molecular clouds (\cite{klein94})). The presence of the baroclinic term
does not change this conclusion and is only important, as we shall
see, when thermal heating is present (\S\ \ref{vishniac}).

\subsection{Other mechanisms of vorticity production}

Since the nonlinear and baroclinic terms have been found in the
previous sections to be globally insufficient for vorticity production
and maintenance in the presence of high compressibility and purely
compressible forcing, it is important to examine alternative sources
of vorticity in astrophysical flows. As an example, we will consider
the case of the ISM at the kpc scale, within which possible agents for
vorticity production are 
the Galactic disk rotation, various heat sources and the magnetic field.

\subsubsection{Coriolis force}\label{coriolis}

It is well known that the Coriolis force can convert converging
motions into shearing ones. In fig.\
\ref{esetlowhi} we also show the time evolution of the ratio $\eset$ for the
three dimensional run labeled run 69, analogous to run 60, but with
the addition of the Coriolis force with $\Omega=0.4$ in code units,
corresponding to the fiducial value $2 \pi /2 \times 10^8$ yr$^{-1}$
(see {\cite{paperII}). It is noticeable that compressible and
rotational modes almost reach equipartition after $4 \times 10^7$
yr. Compared to the run without the Coriolis force (run 60), with the
same random forcing, run 69 develops 30\% less density fluctuations but twice
as much kinetic energy, while the internal energy remains
approximately at the same level. The increase of the level of kinetic
energy is probably due to the fact that compressible motions dissipate
faster than rotational ones and thus, for a given energy input rate, the
equilibrium kinetic energy will be larger if rotational modes are
excited. 
Additionally, it is interesting to estimate the importance of the
Coriolis force at different scales. For economy we do this in
two-dimensions. The Rossby number (measuring the
relative size of the Coriolis force with respect to the nonlinear
advection term), scales as $L^{1/2}$ if we assume that the velocity
dispersion $u$ scales roughly as $L^{1/2}$ (\cite{lar81}; \cite{fpp92}).
A run using $\Omega=0.05$ (run 85), a value appropriate for scales $\sim 25$ 
pc, ends up with only $\sim$ 5\% of
rotational energy, while a run with $\Omega=0.4$ (run 57, appropriate for the
kpc scale) typically contains $\sim$ 50\% of rotational
energy. Interestingly, this value is very similar to the 3D case, run
69. This indicates that the Galactic Coriolis 
force is likely to be an
important source of vortical motions at the kpc scale, but not at the
molecular cloud scale. 

\subsubsection{Large-scale shear from differential rotation}\label{shear}

Another agent due to the Galactic rotation is the shear arising from its
differential character. A run with an imposed shear as performed in
\cite{paperII} corresponding roughly to Oort's A constant (run 59), shows a
selective creation of vorticity of the same sign as that of the imposed
shear. For this run, $\eset \sim 30$\% on average.

\subsubsection{Local heating}\label{vishniac}

The stellar heating $\Gammas$ provides another source of vorticity in
the flow. Indeed, at locations where thermal energy is injected, the
baroclinic term will be important, since the gradients of density and
temperature cease to be correlated. This effect is most noticeable in
the dynamics of expanding shells, which propagate without deformation
when only wind-type forcing  is included in the equations, while they
distort and fragment in the presence of stellar heating. This is
illustrated in fig.\ \ref{expsh},
which shows the density  
fields for runs 28 (left) and 35 (right). These runs start with
purely compressible velocity fluctuations and differ only in
that the forcing in run 35 is in the velocity (through $\nabla \psi$;
cf. \S\ \ref{models}), while run 28 has
thermal energy forcing only (through $\Gammas$). The production of
solenoidal energy is substantial in run 28
in which, during the period of shell breakup, $\eset \sim
0.5$, while in run 35, $\eset$ barely reaches 0.01. Note that the low
value of $\eset$ in run 35 is in agreement
with the result of \S \ref{baroclinic} that the production of vorticity behind
curved shocks is negligible in the absence of external heating.

Note that the breakup of 
shells in the presence of heating can also be interpreted as a consequence of 
linear Vishniac-type instabilities, occurring in fronts between thermal and
ram pressures (\cite{vishniac}). However, the present simulations do
not allow us to decide whether the shells break due to vorticity
production by the baroclinic term, to the linear Visniac instability,
or both. This is because in these runs actually a thin shell of 
star-forming sites forms immediately behind the gas shell, causing the
baroclinic term to be large there. One indication is that in run 35,
with wind-type forcing, the nonlinear Vishniac instability
(\cite{vish94}; see also \cite{hun86}; \cite{sbp92}) does not develop,
suggesting that the present simulations lack the necessary
resolution. A detailed examination of this problem would require
monitoring of the baroclinic term and possibly higher-resolution
simulations. This, however, is beyond the scope of the present
paper. Here we just record the importance of local heating as a source
of vorticity.


\subsubsection{Magnetic field}\label{magnetic}

Finally, we discuss the effect of the Lorentz force on
the transformation of compressive motions into rotational ones. For
this purpose, we have performed four runs with the \ppm\  model labeled 
runs 86, 84, 87
and 88, identical to run 53 except for the inclusion of a uniform
initial magnetic field along the $x$-axis, with strengths of,
respectively, 0.05, 0.3, 1, and 3, in nondimensional code units (at $B_\o=1$,
the Alfv\'en velocity $v_a=u_0$, see Paper II). A 
small fluctuating field with an rms amplitude of 0.01 is also
included. In fig.\ \ref{esetmag} we show the time evolution of the
fraction of specific kinetic energy in solenoidal motions $\eset$ for
the four runs. The fraction of energy in solenoidal modes is seen to
increase with $B_\o$ until equipartition with compressible modes is
reached at values of $B_\o>1$. The
fluctuating magnetic energy obeys the same trend (fig.\ \ref{b2msb0}),
as already mentioned in \cite{paperII} in the context of the full
model. In fig.\  \ref{b2msb0}, temporal oscillations are observed, whose
frequency is roughly proportional to $B_\o$. However, these
oscillations do not appear to be due only to Alfv\'en waves. Instead,
fully developed MHD turbulence is present, as can be
inferred from the presence of locally strong currents -- indicating
small-scale intermittent magnetic structures -- and a power-law magnetic
spectrum (fig.\  \ref{magspectr}).
The increase of the solenoidal energy in the form of fully
developed turbulence with the uniform component of the magnetic field
is remarkable and might, at first sight, appear contradictory to
common assumptions (\cite{elmppIII}, sect.\ VII; \cite{mousch}).
This phenomenon of creation of vorticity by magnetic
fields is important as well in the context of accretion disks (\cite{vidi}).

The relative importance of the
Lorentz force in the momentum equation increases as smaller scales are
considered. For instance, taking typical values of the fluctuating
magnetic field strength of $\sim 5 \mu$G at the kpc scale and $\sim 30
\mu$G at scales $\sim 5$ pc, and typical velocity dispersions of
respectively 10 and 2 km s$^{-1}$, the ratio of the Lorentz force to
the nonlinear advection term is roughly 20 times larger at the 5 pc
scale. Thus, the production of vorticity from compressive motions
due to the Lorentz force is most efficient at molecular cloud scales.

\section{Scale-free flows with self-gravity}\label{selfgrav}

\subsection{Influence of $\game$ on the formation of collapsing
structures} \label{game and collapse}

The results presented in the previous sections have implications for
star formation, since the ability of turbulence to form
self-gravitating clouds depends on the compressibility of the flow. A
simple estimate for the required values of
$\game$ for this to occur can be given as follows. Consider a
medium with an effective
polytropic exponent $\game$, for which the
effective Jeans length is given by 
\begin{equation}
L_{\rm eff} = \Bigl[{\gameff \pi
c_{\rm i}^2 \over G\rho_\o^{2-\game}} \Bigr]^{1/2} = \sqrt{\frac{\game}{\gamma}}
\rho^{\frac{\game-1}{2}} L_{\rm J}, \label{jeans length}
\end{equation}
where $c_{\rm i}$ is the isothermal sound speed such that $P=c_{\rm i}^2
\rho$ and $L_{\rm J}$ is the Jeans 
length based on $c_{\rm i}$. 
The critical density
at which the scale $L$ becomes unstable is then given by $\rho_{\rm J}
\propto L^{2/(\game - 2)}$. 
In order to account for turbulent compressions acting on $n$ directions,
we consider a volume $V=L^n L_0^{3-n}$, where $L$ is the side of the
volume which varies upon compression, and $L_0$ is the side that
remains unaltered. The critical mass to destabilize this volume is
thus given by
\begin{equation}
M_{\rm J} \propto L^{n+\frac{2}{\game-2}} L_0^{3-n}.\label{criterion}
\end{equation}
If $M_{\rm J}$ is a decreasing function of $L$,
turbulent compression can produce gravitationally unstable
structures, similarly to Hoyle's (1953) notion of hierarchical
gravitational fragmentation. Setting the exponent of $L$ to zero then 
gives the critical value of the polytropic exponent
as a function of the dimensionality of the compression $n$. In
particular, 
\begin{equation}
\gamma_{\rm c} = 
\cases{\frac{4}{3}      & for $n=3$\cr
        1               & for $n=2$\cr
        0               & for $n=1$.\cr
}
\end{equation}
This result thus recovers the well-known critical values of $\game$
for the existence of equilibrium solutions of self-gravitating
polytropic gas configurations (e.g., \cite{chandrabook}), although within
the different context of $n$-dimensional turbulent compressions. In
the present case, this criterion describes the threshold value $\gamma_c$ of
$\game$ at which the internal energy of a turbulent fluctuation
increases more slowly upon compression than its gravitational
energy. Conversely, above $\gamma_{\rm c}$, a fluctuation can never be
rendered unstable upon compression. Note
that for $\game < 0$, no compression is necessary because the medium is then
thermally unstable. 

If we extend this result to non-integer values of $n$, accounting for
clouds with fractal boundaries of typical dimension
$1.4$ (\cite{fpw91}), and recalling that the
compression is likely to be driven by quasi-planar shocks (see also
\cite{fle96}),
we obtain that $0 < \gamma_{\rm c} \leq 1$, which is probably not
unrealistic.

In order to examine the interplay between turbulence, compressibility
and self-gravity, we have
performed a series of forced runs with the B-model, using barotropic
exponents of 0.9, 0.3 and 0.1 (runs 106, 107 and 108 respectively) and
the same Jeans length $L_{\rm eff}=2 \pi /J$ (which is independent of
$\game$, a fact following from the
form of the pressure for the B-model), with $J=0.9$. Note that all
these runs are stable according to the linear criterion. 
The evolution
of the maximum density within the integration box for runs 106, 107,
and 108 is shown in fig.\ \ref{densevol 106-108}. The turbulent
density fluctuations never grow in the case of the high-gamma run, but
in the case of runs 107 and 108, local fluctuations eventually collapse,
slightly earlier in the latter run, confirming earlier speculations by
various workers (e.g., \cite{hunfle82};
\cite{bonaz}; \cite{leorat}; \cite{padoan}) and extending the results of Hunter
et al.\ (1986) for isolated gas streams to the fully turbulent regime.
In fact, given the identical initial conditions and random forcing
used in the two runs, the condensation that collapses in run 108 also
forms in run 107, but is unable to actually collapse. 

It is important to emphasize that
the collapse in the turbulence-induced fluctuations is at small scale,
contrary to the type of collapse expected for the linearly unstable
case.
To show this, we have performed an additional forced run with $J=1.1$
and $\game=0.9$ (run 109). 
Contours of the logarithm of the 
density field for runs 109 and 108 are presented in fig.\ 
\ref{cont108}, in which the left frame corresponds to run 109 and the
right frame to run 108. The large void around the
condensation in run 109 shows that most of the mass in the
simulation is involved in the collapse, while significant amounts of
matter are present outside the condensation for run 108. Also, note
that the ``stable'' run (run 108) takes a rather long time ($t=14$) 
to finally develop a
collapsing cloud, since a strong enough turbulent fluctuation is
required, which is statistically unlikely. 
This may be an
appropriate mechanism behind the low efficiency of star formation
observed in the ISM (see, e.g., \cite{evans91}).
Instead, run 109, which is
globally unstable, collapses at $t=8.6$, a time consistent,
within a factor of 1.5,
with the free-fall time $t_{\rm ff}\equiv L_{\rm J}/c =2 \pi/(J \rho) =
5.7$ in code units.

Finally, we note that the runs described in this section, although
providing strong support for the validity of the criterion
(\ref{criterion}), cannot be used to determine the actual value of
$\gamma_{\rm c}$, since in order to do this it would be necessary to
verify that arbitrarily large density fluctuations do not collapse,
which for large $\game$ would require very large Mach
numbers. Moreover, the dimensionality $n$ of individual shocks may vary,
and therefore the value of $\gamma_{\rm c}$ is probably meaningful only in a
statistical sense.

\subsection{Magnetic field and self-gravity} \label{mhdself}

We now investigate whether this mechanism is also at play in the
presence of a uniform magnetic field $B_\o$, which we take along the
$x$-direction. The instability criterion of the uniform-density state
is unchanged compared to the non-magnetic case (which 
reads $J > 1$), except for perturbations exactly perpendicular to the
field, for which the
corresponding condition reads $J^2 > 1+ B_\o^2$ (\cite{chandrabook}).
%
%
Again, note the independence on $\game$ in these criteria in the context of the
B-model. We have performed a forced run similar to run 109 but with $B_\o =2$
in code units (run 111). This run, with $J=1.1$ and $\game=0.9$, 
undergoes contraction along the field lines,
but the resulting slab remains stable in the direction tranverse to
the field. Note that the critical value $B_1$ of the uniform magnetic 
field for linear stability of
the slab should be in fact larger than that corresponding to the
transverse stability of the uniform state.
A similar run with $\game=0.3$ (run 114) also forms a slab, but the
turbulence is able to create a subcondensation within the slab which
collapses gravitationally, thus exhibiting the same qualitative behavior as
that reported in the previous subsection for the non-magnetic
case. This is illustrated in fig.\ \ref{slab}, which shows contour
plots of the density for both runs.

The case described above suggests that the system has a supercritical 
mass to flux ratio (\cite{mousch&spit}), since
the magnetic field is not able to stop the collapse. This behavior is
still observed with $B_\o=2.5$, but at $B_\o=3$ the system again does
not collapse in spite of the production of fluctuations
exceeding $\rho=55$, suggesting that  the system has
become subcritical. We performed new runs with the B-model, 
no forcing and $\nu_2=\eta=0.04$, taking as  initial conditions zero
velocity dispersion and the mass concentrated in a wide
Gaussian profile at the center of the box with $\gamma_e=0.9$, a choice of
parameter that permits to avoid violent initial collapse transients.
For both $B_\o=3$ and $B_\o=4$ the system reaches an
equilibrium. Starting now new simulations from these equilibrium states
with $\game=0.3$ leads, for $B_\o=4$, to a new equilibrium with a
higher density peak, while for $B_\o=3$ the system collapses,
indicating that for that value of $B_\o$, thermal pressure is still
needed to fight gravity.  As a result we can argue that an arbitrary
high level of turbulence would not be able to produce gravitational
collapse for $B_\o=4$, whereas if a strong enough fluctuation could
gather enough mass, collapse could occur with $B_\o=3$.

Thus, at a fixed value of $J$, 
there are two critical values of the uniform field $B_\o$: one ($B_1$) to 
provide linear
stability of the slab state, and the other ($B_2>B_1$)
corresponding to the
transition from supercritical to subcritical. According to our
results, turbulence is able to shift $B_1$ to higher values as
turbulent fluctuations lose thermal support (at small enough $\game$). 
However, it does not seem
to be able to modify $B_2$. We speculate that $B_1$ should tend to
$B_2$ as thermal support is weakened ($\game \rightarrow 0$).
Therefore, the mass-to-flux ratio criterion appears
to be very robust to the presence of strong nonlinearities (at the large
scale),
as it is based solely on global conservation
conditions. However, we note that the presence of
small-scale MHD turbulence inside the condensations themselves, which we cannot
follow here due to the limited resolution, could in principle
provide, through reconnection, a dissipation of the magnetic field
allowing for an increase of the mass-to-flux ratio. Finally, we remark
that, although turbulence appears incapable of inducing collapse of
supercritical condensations, it is a necessary agent in their formation,
since the slab state is linearly stable in the absence of turbulence
due to the contribution from the thermal pressure.

\section{Behavior of a thermodynamic model of the ISM at the kpc
scale} \label{allinc}

In this section we test the applicability of our conclusions on the
influence of $\game$ on cloud formation and collapse scenarii in
the context of the full model used in \cite{paperII} (cf.\ \S\
\ref{models}) including also rotation and shear, with values
adequate for the ISM at the 1 kpc scale. We have performed two 
simulations (runs 102 and 113) with the fiducial
values described in Table 1 of \cite{paperII}, although without star formation 
nor random forcing ($\Gammas=0, {\bf F}_c=0$). These two runs 
differ only in the value of $\alpha$. We do not include forcing
since in this case it should be at small scales due to stellar
activity, which in turn should be the outcome of gravitational collapse.

Neither run 102 nor run 113 collapse gravitationally, at least over
50 code time units. (Recall that for this model,
a code time unit corresponds to $1.3 \times 10^7$ yr and a density unit
equals 1 cm$^{-3}$.)
Since no forcing other than the large scale shear is present in runs
102 and 113, the turbulence decays after a few turnover times and we
only observe the nonlinear development of the oscillations suggested
by Elmegreen (1991) and described in \cite{paperII}. As in \cite{paperII},
the clouds form preferentially in regions of zero shear (since shear
rotates a perturbation contracting along the field into the
perpendicular direction); other runs
without magnetic field (not shown) form condensations of even lower
amplitude (due to the absence of magnetic braking)
and in regions of minimum shear, i.e. a shear which minimizes the epicyclic
frequency. The absence of runaway collapse in these runs is interesting
since, according to the linear criterion of \cite{paperII}, they should
be unstable. This confirms the speculation by Elmegreen (1991) and the
observation of \cite{paperII} that the development of the instability
is not exponential, but oscillatory, although in the present paper the
density fluctuations have reached much larger amplitudes, and their
ability to rebound is remarkable. Ultimately, the possibility that
they will collapse on very large time scales cannot be ruled out, but is not
necessarily relevant in the context of the ISM.

In spite of the
absence of gravitational collapse, the simulations support the
scenario of the previous sections: the hard case ($\alpha=0$)
develops only one
condensation of fluctuating amplitude, reaching $\rho_{\rm max} =
20$ at $t=14$ in code units. 
Instead, the $\alpha=0.5$ case develops a larger number of
condensations, one reaching $\rho_{\rm max}= 86$ at $t=7.6$, which
however rebounds and does not collapse. Later, this same condensation
reaches $\rho_{\rm max}=95$ at $t=20.6 $. The evolution of the maximum
density in the integration box is shown for both runs in fig.\
\ref{evol 102 113}. The eventual dying out of such fluctuations as seen
here for late times is linked to the choice $\Gammas=0$, as discussed in 
Paper I and II. These condensations move along
epicycles, and exhibit a wave-like behavior since there are no
turbulent compressions nor instabilities of any kind (\S\
\ref{denstruct}; see also
\cite{adams94}; \cite{gehman96}). 

The latter observation is a consequence of the well-known fact that
the presence of rotation restores a ``hard'' behavior, analogous to
a large value of $\game$. However, as discussed in \S\ \ref{coriolis},
the relative importance of the Coriolis force compared to the magnetic
and nonlinear advection terms decreases at small scales, so that this
scenario suggests stability of the large scales with possible
instability of the small scales. Moreover, the turbulence-induced
collapse of the small scales shown in \S\ \ref{selfgrav} should also
contribute to small-scale collapse, although in particular in runs 102
and 113 this effect cannot be seen due to the absence of fully
developed turbulence and to the low resolution. 

\section{Conclusions} \label{concl}

In the present paper, we have investigated some properties of highly
compressible turbulence, which is relevant in flows with very efficient
cooling processes, such as the ISM or cosmological flows after the epoch
of recombination. In these
cases, the flow may  be subsonic with respect to the isothermal sound
speed, and yet have an effectively supersonic behavior.
We first considered the purely hydrodynamical regime, addressing 
the type of density structures formed by turbulence. We
showed that the density jump in barotropic flows in the limit of small
polytropic index $\game$ approaches $e^{M^2}$, where $M$ is the Mach
number ahead of the shock, rendering the flow much
more compressible than high Mach number regimes with $\game \sim 1$. 
When $\game$ is small, the density structures are long-lived and
object-like, i.e. carrying the mass together with the density
fluctuations while, for larger values of $\game$, the behavior is
more wave-like, and the density field consists of short-lived patches
separated by shocks.

In the presence of purely compressible forcing, 
vorticity is observed to decay in time, even in 3D,
where the dominant effect of the vortex stretching term is actually to
act as a sink of potential vorticity at the location of shocks. Thus
vorticity production relies heavily on other sources, such as the
Coriolis force, large-scale shear originating from differential
rotation, baroclinic vorticity production in the presence of
local heating, and the magnetic field. For the Galactic
disk, the Coriolis force is most important at large scales, while the
magnetic field is most effective at the small ones (molecular clouds).
The decay of vorticity we observe in the absence of those additional agents
may be at the origin of the small amounts of rotation found in
observational surveys of molecular clouds and their cores (e.g.,
\cite{arqgol86}; \cite{good93}).

With self-gravity, we have given a simple criterion for turbulent
compression through shocks to produce
self-gravitating structures, based on the threshold value of $\game$
below which the Jeans mass of a turbulent density fluctuation
decreases upon the compression. 
In particular, if the dimensionality of
the compression is between 1 and 2, then the required value of
$\game$ is between 0 and 1, consistent with estimated values in
the ISM (e.g., \cite{myers78}). Using numerical simulations, we have
verified that in the presence of strongly compressible turbulence,
local turbulent fluctuations can collapse even if the medium is
globally stable in the linear sense, although generally requiring long
times. Flows with small $\game$ develop collapsing
structures at smaller scales than flows with larger $\game$ but with
the same effective Jeans length. Flows with self-gravity and 
magnetic fields exhibit an analogous
behavior to the purely hydrodynamic case for supercritical
mass-to-flux ratios. For subcritical situations, large-scale turbulence is
incapable of inducing small-scale collapse, although it is a necessary
agent for the formation of magnetically supported clouds.

These results have implications for
present-day {\it vs.} primordial star formation, inasmuch as $\game$
is directly related to the medium's metallicity. For the low metallicity
primordial gas, 
we expect that turbulence will not be sufficient to produce
small-scale collapsing structures due to the large values of $\game$,
and therefore collapse can only occur through large-scale
gravitational instability, leading to the formation of globular
clusters. 
Conversely, collapsing turbulence-induced, small-scale density
fluctuations may lead to formation of open clusters under
present-day metallicity conditions. This problem will be investigated in detail
in a forthcoming paper (\cite{franco}).

Fully thermodynamic models incorporating magnetic fields, rotation and
shear with parameters appropriate to the modeling of the ISM at the
kpc scale (c.f.\ \cite{paperII}) but without stellar heating ($\Gammas=0$)
develop highly nonlinear wave-like
structures which however do not collapse gravitationally. Their
behavior is analogous to the hydrodynamic case with a larger $\game$
and initial turbulent transients are not capable of forming small-scale
self-gravitating structures. Thus, in order to obtain collapse of the
large scales an instability is required. 
On the other hand, small-scale collapse in the clouds
should also occur, since rotation is less efficient at those scales
and we have shown that turbulence also induces it. This is not
observable in the simulations presented in this paper due to their low
resolution, but may be seen at higher resolutions.

A final remark is that in this paper we have investigated the effects
of large-scale turbulence on structure formation in highly
compressible flows, but have neglected small-scale effects, such as
those mentioned above, as well as turbulent pressure support and
reconnection of small-scale
magnetic fluctuations, which could
possibly modify the mass-to-flux ratio and
allow for an increased compressibility of the flow. The simultaneous
investigation of both aspects of turbulence requires very high
resolution simulations, which will be presented elsewhere.

\begin{acknowledgements}

We are glad to acknowledge John Scalo for useful comments and
suggestions, and a critical reading of the manuscript. 
The numerical simulations were performed on the Cray C98 of IDRIS,
CNRS, France, and the Cray Y-MP 4/64 of DGSCA, UNAM, M\'exico. This work has
made use of NASA's Astrophysical Data System Abstract Service. We
thankfully acknowledge financial support from UNAM/CRAY grant SC-002395 to
E.\ V.-S., and a joint CONACYT-CNRS grant. 

\end{acknowledgements}

\vfill\eject

\begin{figure}[htb]
\plotone{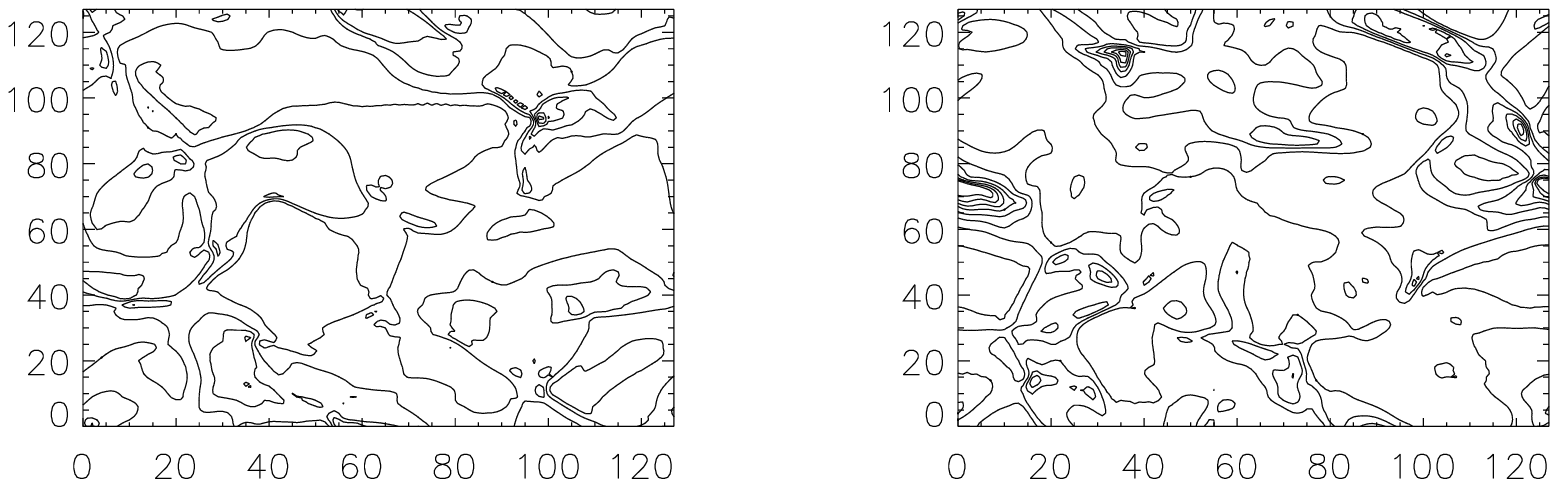}
\caption{Contours of the density field for run 63 ($\game=1.5$, {\it left})
    and run 65 ($\game=0.3$, {\it right}) at a nondimensional  
    time of $t=7.6$. In both cases
    there are 10 contour levels with a
    constant spacing $\Delta \rho =0.3$. Note the ``patchy'' structure
    of the density field, with higher peaks at the intersection of
    shocks in (b).}
\label{hilowgam}
\end{figure}

\begin{figure}[htb]
\plotone{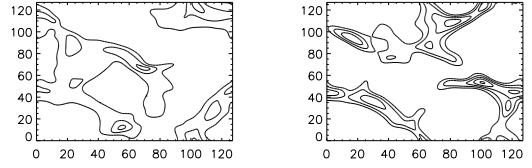}
\caption{Contours of the density field of run 54 ($\alpha=-0.5$, 
  {\it left}) and
  run 53 ($\alpha =0.5$, {\it right}) at
  $t=6.0$. The contours represent density values of
  $\rho=1,2,4,8,16,32$. Both runs use the \ppm\ equations. Note that
  clouds are better defined than in fig.\ \protect \ref{hilowgam},
  and have higher peak densities in (b).}   
\label{ppmhilow}
\end{figure}

\begin{figure}[htb]
\plotone{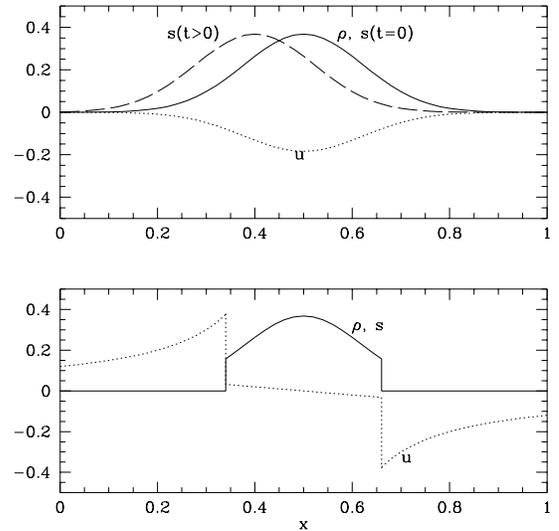}
\caption{One-dimensional sketches of the density $\rho$, the passive
  scalar $s$ and the velocity $u$ in the case of a wave ({\it top}) and an
  object ({\it bottom}). In the case of the wave, the passive scalar,
  which represents advection of material, drifts away from the
  density, while in the case of the object, $s$ remains within the
  density fluctuation.}
\label{rhov phases}
\end{figure}

\begin{figure}[htb]
\plotone{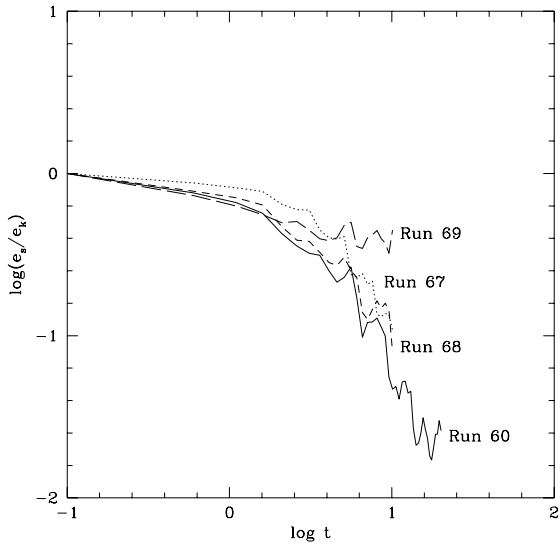}
\caption{Evolution of the ratio of solenoidal to total kinetic energy
  per unit mass $\eset$ for runs 60 ({\it solid line}), run 67 ({\it dotted
    line}), run 68 ({\it short-dashed line}) and run 69 ({\it
    long-dashed line}). Runs 60 and 67 are in 3D and 
    use the \ppm\ equations with
  $\alpha= 0.5$ and $\alpha=-0.5$ respectively. Run 68 is similar to
  run 60 but fully thermodynamic. Run 69 is similar to run 60 but
  includes the Coriolis force with $\Omega =0.4$. After similar initial
  transients, the ratio $\eset$
  decays in all cases except in the run including the Coriolis force.}
\label{esetlowhi}
\end{figure}

\begin{figure}[htb]
\plotone{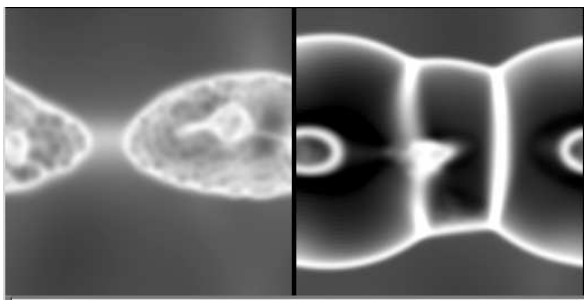}
\caption{Gray-scale images of the density field for run 28 ({\it left}) 
    and run 35 ({\it
    right}). Both runs are fully thermodynamic, but run 28 has local thermal
  forcing only, while run 35 has wind-type
  compressible forcing on the velocity only. 
  The shells evolve smoothly in run 35, but distort and
  disrupt in run 28 due to the strong vorticity production.}
\label{expsh}
\end{figure}

\begin{figure}[htb]
\plotone{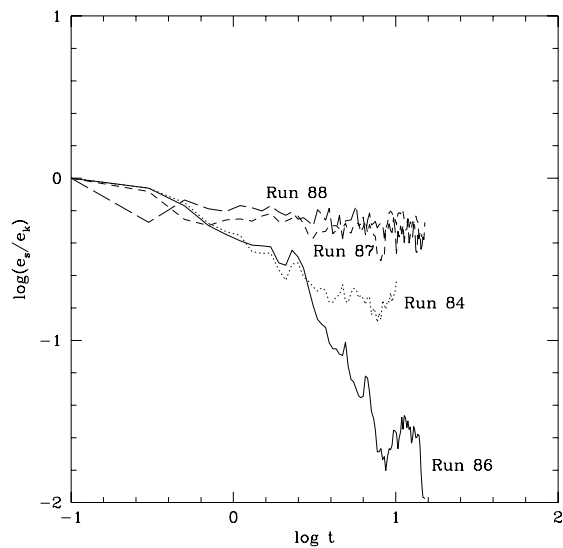}
\caption{Evolution of the ratio of solenoidal to total kinetic energy
  per unit mass $\eset$ for runs 86 ({\it solid line}), run 84 ({\it dotted
    line}), run 87 ({\it short-dashed line}) and run 88 ({\it
    long-dashed line}). These runs respectively have values of the
  initial uniform field of $B_\o=0.05, 0.3, 1$ and 3. The final value
  of $\eset$ is seen to increase with $B_\o$ until saturation at the 
  equipartition level is reached for $B_\o>1$.}
\label{esetmag}
\end{figure}

\begin{figure}[htb]
\plotone{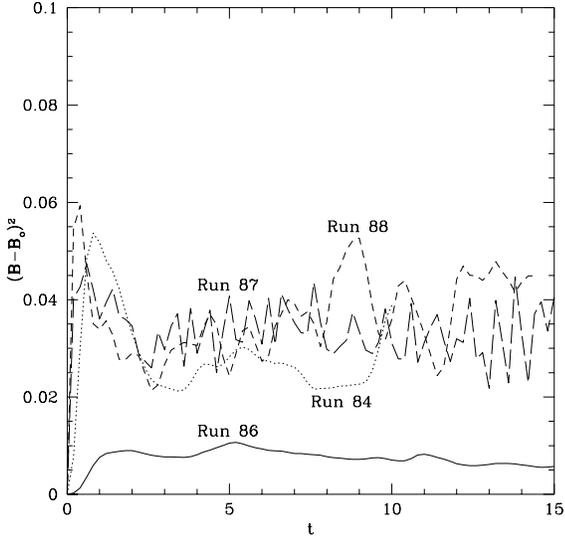}
\caption{Evolution of the fluctuating magnetic energy for the same
  runs as in fig.\ \protect \ref{esetmag}, using the same line
  coding. A similar trend of increasing strength with $B_\o$ is
  observed, with saturation at $B_\o>1$. Note also the
  increasing frequency of the oscillations, roughly proportional to $B_\o$.}
\label{b2msb0}
\end{figure}

\begin{figure}[htb]
\plotone{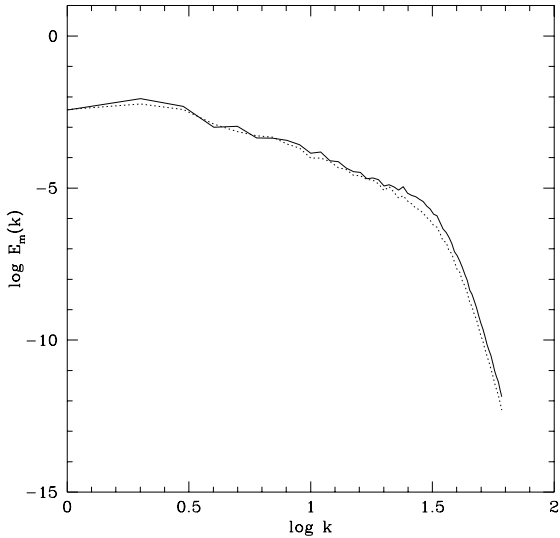}
\caption{Spectra of the $x$- and $y$-components of the
  magnetic field of run 87 at $t=14.4$. A well-defined power-law
  inertial range is observed, indicative of fully developed turbulence.}
\label{magspectr}
\end{figure}

\begin{figure}[htb]
\plotone{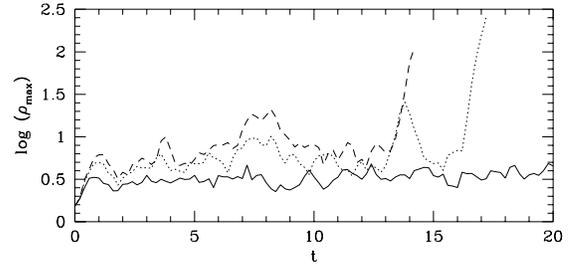}
\caption{Evolution of the maximum density for runs 106 ({\it solid
    line}), 107 ({\it dotted line}) and 108 ({\it dashed line}),
  which respectively have $\game=0.9, 0.3$ and 0.1, all with $J=0.9$,
  implying linear stability. Turbulent fluctuations of finite strength
  are seen to be
  able to collapse gravitationally for runs 107 and 108 in spite of
  the global linear stability.}
\label{densevol 106-108}
\end{figure}

\begin{figure}[htb]
\plotone{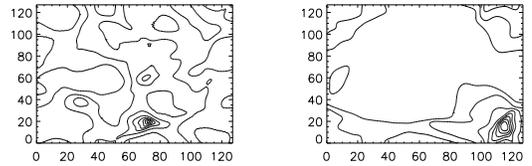}
\caption{Contour plots of the density fields of runs 108 ({\it left})
  and 109 ({\it right}), which
  respectively have $J=0.9$ and 1.1. Run 109 is thus linearly
  unstable, and the collapse is seen to involve most of the mass in
  the simulation, while the turbulence-induced collapse in run 108 is
  local. The contours denote the values $\rho=0.5,1,2,4,8,16,32$ and 64.}
\label{cont108}
\end{figure}

\begin{figure}[htb]
\plotone{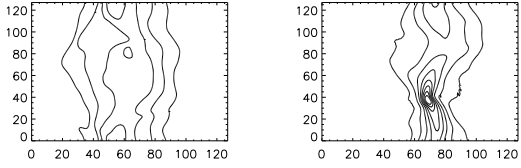}
\caption{Contour plots of the density fields of runs 111 ({\it left})
  and 114 ({\it right}), both
  permeated by a uniform magnetic field of
  strength $B_\o=0.2$ in the $x$-direction and with $J=1.1$. Run 111 has
  $\game=0.9$, and for run 114, $\game=0.3$. Being linearly unstable,
  both runs contract to a slab. Turbulent fluctuations, however,
  manage to produce a subcondensation in run 114 which finally
  collapses gravitationally. The contour spacing is identical to that of
  fig.\ \protect \ref{cont108}.}
\label{slab}
\end{figure}

\begin{figure}[htb]
\plotone{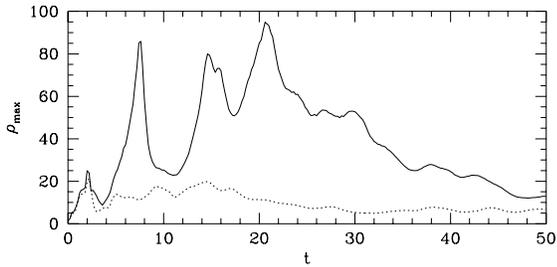}
\caption{Evolution of the maximum density for runs 102 ({\it solid
    line}) and run 113 ({\it dotted line}). Both runs include
  rotation, shear and magnetic field, in the context of the full model
  of Paper II, but without star formation. Although not collapsing
  gravitationally, both runs exhibit strong density fluctuations.}
\label{evol 102 113}
\end{figure}

\clearpage
\vfill\eject

\brokenpenalty=10000
\centerline{{\bf Table 1.} Run parameters}
\bigskip
\begin{tabular}{llllllllll}
\hline
Run $\#$ & Type$^a$ & Forcing$^b$ & $(e_s/e_k)_0^{\ c}$ & Shear$^d$ & $J^e$ 
& $B_0^{\ f}$ & $\Omega^g$ & $\gamma_e^{\ h}$ & $\alpha^i$ \\
\hline
28& T2D & W & 0 & N & 0 & 0 & 0.0 & N/A & 0.5\\
35& T2D & T & 0 & N & 0 & 0 & 0.0 & N/A & 0.5\\
53& ppm2D & R & 1 & N & 0 & 0 & 0 & N/A & 0.5\\
54& ppm2D & R & 1 & N & 0 & 0 & 0 & N/A & -0.5\\
57& ppm2D & R & 1 & N & 0 & 0 & 0.4 & N/A & 0.5\\
59& ppm2D & R & 1 & Y & 0 & 0 & 0.0 & N/A & 0.5\\
60& ppm3D & R & 1 & N & 0 & 0 & 0 & N/A & 0.5\\
63& B2D & R & 1 & N & 0 & 0 & 0 & 1.5 & N/A\\
65& B2D & R & 1 & N & 0 & 0 & 0 & .3 & N/A\\
67& ppm3D & R & 1 & N & 0 & 0 & 0 & N/A & -0.5\\
68& T3D & R & 1 & N & 0 & 0 & 0 & N/A & 0.5\\
69& ppm3D & R & 1 & N & 0 & 0 & 0.4 & N/A & 0.5\\
84& ppm2D & R & 1 & N & 0 & 0.3 & 0 & N/A & 0.5\\
85& ppm2D & R & 1 & N & 0 & 0 & 0.05 & N/A & 0.5\\
86& ppm2D & R & 1 & N & 0 & 0.05 & 0 & N/A & 0.5\\
87& ppm2D & R & 1 & N & 0 & 1. & 0 & N/A & 0.5\\
88& ppm2D & R & 1 & N & 0 & 3. & 0 & N/A & 0.5\\
102& T2D & N/A & 1 & Y & .51 & .3 & 0.4 & N/A & 0.5 \\
106& B2D & R & 1 & N & 0.9 & 0.0 & 0 & 0.9 & N/A\\
107& B2D & R & 1 & N & 0.9 & 0.0 & 0 & 0.3 & N/A\\
108& B2D & R & 1 & N & 0.9 & 0.0 & 0 & 0.1 & N/A\\
109& B2D & R & 1 & N & 1.1 & 0.0 & 0 & 0.9 & N/A\\
111& B2D & R & 1 & N & 1.1 & 2.0 & 0 & 0.9 & N/A\\
113& T2D & N/A & 1 & Y & .51 & .3 & 0.4 & N/A & 0. \\
114& B2D & R & 1 & N & 1.1 & 2.0 & 0 & 0.3 & N/A\\
\hline
\end{tabular}

\clearpage
\vfill\eject
\centerline{\bf Table 1}

Main characteristics of the runs described in this paper, together with 
the run number. All runs except run 35 exclude stellar heating
(i.e. with $\Gammas=0$ in (\ref{ener})).
Computations are done on a uniform grid of either $128^2$ or $64^3$ points, 
using a pseudo-spectral method with hyperviscosity. For more detail on
the implementation of the physical terms and in the nondimensionalization
of the equations, see text and Paper I and II.
\vskip0.08truein

\par\noindent $^a$ \hskip0.09truein Types of computations:
{\sl T2D} (resp. {\sl T3D}) refers to the full thermodynamical equations 
(\ref{cont}) -- (\ref{lamda}) written in \S 2, integrated in two space 
dimensions (resp. three);
{\sl ppm2D} (resp. {\sl ppm3D}) refer to a simplification of the full equations
in which a {\it piecewise polytropic model} (see (\ref{pppm})) is used
in two (resp. three) dimensions; and finally {\sl B2D} (resp. {\sl B3D})
correspond to the further simplification of a barotropic law in two 
(resp. three) dimensions.
\vskip0.04truein

\par\noindent $^b$ \hskip0.09truein The type of forcing used in the different 
models can be either:
(R), a random forcing with a steep ($\sim k^{-4}$) spectrum;
(T), a thermal energy forcing with $\Gammas \not= 0$;
or finally (W), a wind with ${\bf F}_c=-\nabla \psi$ with $\psi \sim \Gammas$.
\vskip0.04truein

\par\noindent $^c$ \hskip0.09truein $(e_s/e_k)_0$ is the ratio at $t=0$ of 
the rotational to total kinetic energy.
\vskip0.04truein

\par\noindent $^d$ \hskip0.09truein Y (resp. N) indicates the presence (resp.
absence) of shear in the computation.
\vskip0.04truein

\par\noindent $^e$ \hskip0.09truein $J$ is the gravitational parameter and 
measures the number of Jeans masses in the flow; when $J=0$, 
no self-gravity is included.
\vskip0.04truein

\par\noindent $^f$ \hskip0.09truein $B_\o$ is the mean magnetic field;
$B_\o=1$ corresponds to an Alfv\'en velocity equal to the rms turbulent 
velocity (see Paper II). In this column, $B_\o\equiv 0$ indicates a
non-magnetic run.
\vskip0.04truein

\par\noindent $^g$ \hskip0.09truein $\Omega$ measures the intensity of the
Coriolis force, independently of the externally imposed shear.
\vskip0.04truein

\par\noindent $^h$ \hskip0.09truein $\gamma_e$ is the effective polytropic
exponent used in the purely barotropic (B) models (see text).
\vskip0.04truein

\par\noindent $^i$ \hskip0.09truein $\alpha$ is the exponent of the
density used 
in a model of diffuse heating (see Paper II). The relationship between
$\alpha$ and ${\gameff}_{i}$, the set of effective polytropic exponents
appearing in the \ppm\  models is given in eq.\ (\ref{albega}).

\end{document}